\documentclass[,final]
  {aipproc}

\layoutstyle{8x11double}

\begin{document}

\title{Craters Formed in Granular Beds by Impinging Jets of Gas}


\author{Philip T. Metzger}{
  address={Granular Mechanics and Regolith Operations Laboratory, NASA Kennedy Space Center, Florida 32899, USA}
}

\author{Robert C. Latta, III}{
  address={Department of Aerospace Engineering, Embry-Riddle Aeronautical University, Daytona Beach, Florida 32114, USA}
}

\author{Jason M. Schuler}{
  address={ASRC Aerospace, Kennedy Space Center, Florida 32899, USA}
}

\author{Christopher D. Immer}{
  address={ASRC Aerospace, Kennedy Space Center, Florida 32899, USA}
}

\begin{abstract}
When a jet of gas impinges vertically on a granular bed and forms a crater, the grains may be moved by several different mechanisms:  viscous erosion, diffused gas eruption, bearing capacity failure, and/or diffusion-driven shearing.  The relative importance of these mechanisms depends upon the flow regime of the gas, the mechanical state of the granular material, and other physical parameters.  Here we report research in two specific regimes: viscous erosion forming scour holes as a function of particle size and gravity; and bearing capacity failure forming deep transient craters as a function of soil compaction.
\end{abstract}

\maketitle

\section{Introduction}

Jets of fluid impinging vertically on granular beds are scientifically interesting due to the variety of phenomena that have been observed.  The present study was motivated by a desire to understand how landing rockets will affect the soil of an extraterrestrial body.  Small-scale experiments help to identify basic relationships in the physics and provide a dataset to benchmark the high-fidelity numerical flow codes that are being developed.

So far researchers have identified four basic ways that an impinging jet of gas (or liquid) can cause a crater to form in a granular bed.  First, \textit{viscous erosion} (VE) is the process of fluid lifting or rolling the top layer of grains along the surface \cite{bagnold}, forming a scour-hole in the case of a localized jet.  Second, \textit{diffused gas eruption} (DGE) occurs when the jet pressurizes the interstitial fluid of the granular bed in a radially-expanding subsurface region, which may cause an eruption of granular material in an annular ring around the jet, or centrally if the jet is suddenly extinguished \cite{scott}.  Third, \textit{bearing capacity failure} (BCF) occurs when the dynamic pressure of the impinging jet applies mechanical loading to the top surface of the granular bed beyond its bearing capacity and thus shoves it downward to form a depression \cite{alexander}.  Fourth, \textit{diffusion-driven flow} (DDF) occurs when the fluid driven by a jet through the pore spaces of the soil creates a distributed body force in the bulk of the soil---the drag force of the fluid reacting against the grains---sufficient to unjam the material and shear it \cite{metzger2}.  The shearing of DDF is geometrically different than BCF, which is caused by forces acting on the free surface of the bed.  If the fluid diffusing through the soil reaches steady-state faster than the soil can shear, then BCF is impossible and only DDF can occur.  However, if the sudden application of a jet onto the granular bed produces enough surface load to shear the grains faster than the fluid can diffuse into the pore spaces, then DDF is impossible and only BCF can occur.  In general the cratering is intermediate to these two extremes. Experiments have shown that BCF causes the grains to move perpendicularly away from the surface, whereas DDF causes them to move parallel to the surface, and in the intermediate cases the grains move diagonally away from the surface \cite{metzger2}.

It would be scientifically interesting and important to a wide range of engineering applications to map the parameter space of these four jet-induced cratering mechanisms to know where each becomes important and how they may interact.  To-date this has not been attempted; research has only sampled the various behaviors at some convenient points in parameter space.  Here we report two such samplings.  We report on the scaling of VE forming scour holes beneath subsonic jets with variation in the granular particle size and gravity, and we report on the insensitivity of BCF to soil properties beneath supersonic jets.

\section{Scour Holes beneath Subsonic Jets}

\begin{figure}
  \includegraphics[width=\columnwidth]{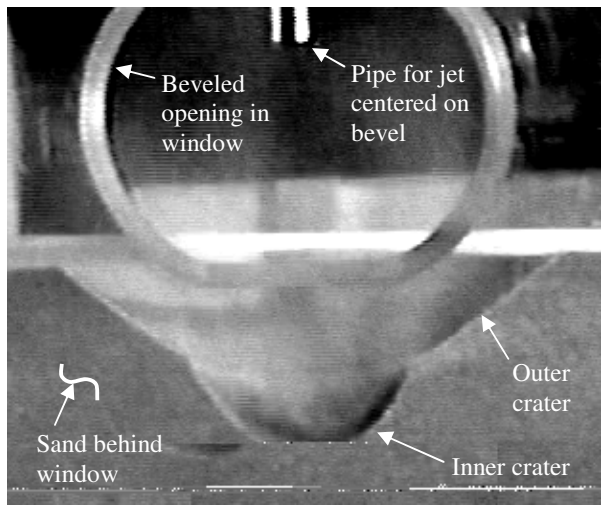}
  \caption{\label{cratershape} Inner and outer craters beneath a subsonic jet.}
\end{figure}

The depth of the scour holes formed by VE under subsonic jets was reported by Rajaratnam and Beltaos \cite{rajbelt} to grow logarithmically over several decades of time (what we shall call the ``logarithmic period'') before slowing to reach an asymptotic depth.  Prior studies have investigated the asymptotic size and shape of these scour holes \cite{rajbelt, raj, ansan} and the transient growth of the holes \cite{metzger2, haehnel1}, but much of the parameter space is yet to be explored.  Here we have used the experimental method reported in detail in \cite{metzger2}, directing a jet of nitrogen gas from a straight circular pipe into a small sand box so that it forms a crater.  The controlled parameters are jet velocity, height and diameter of the pipe above the sand, and sand composition.  The experiment is effectively split in half by replacing the front wall of the sand box with a viewing window and centering the circular pipe over its outwardly beveled edge.  Thus we can see inside the crater as it forms, as shown in Fig.~\ref{cratershape}.  The crater's inner surface is found to become steeper than the angle of repose because it is supported by the traction of the gas exiting the crater.  Eventually the crater widens and traction is insufficient to maintain the steep slope, so the upper part of the crater collapses to form an ``outer crater'' at the angle of repose as shown in Fig.~\ref{cratershape}
We video-recorded the craters through the viewing window and measured their size and shape at each video frame using automated software.  From these measurements we extracted their scaling behavior.  

\subsection{Scaling With Particle Size}

\begin{figure}
  \includegraphics[width=\columnwidth]{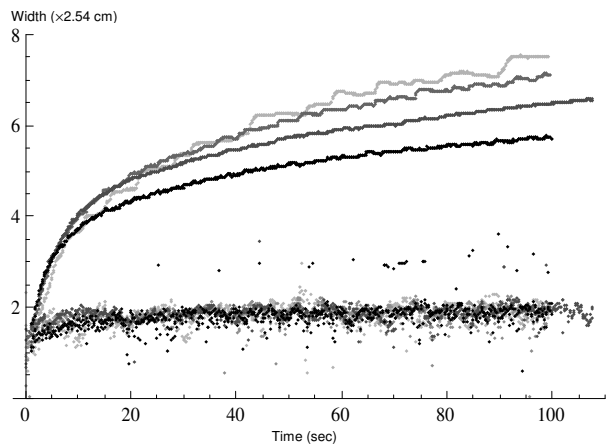}
  \caption{\label{craterwidths} Inner and outer craters widths versus time for four sand sizes (in microns) 200-280 (lightest gray), 280-300, 300-450, and 500-600 (darkest gray).  The top set of curves shows outer crater widths.  The bottom set of curves (overlying one another) shows inner crater widths.  The noisiness of the inner craters is caused by avalanching of the outer craters.}
\end{figure}

\begin{figure}
  \includegraphics[width=\columnwidth]{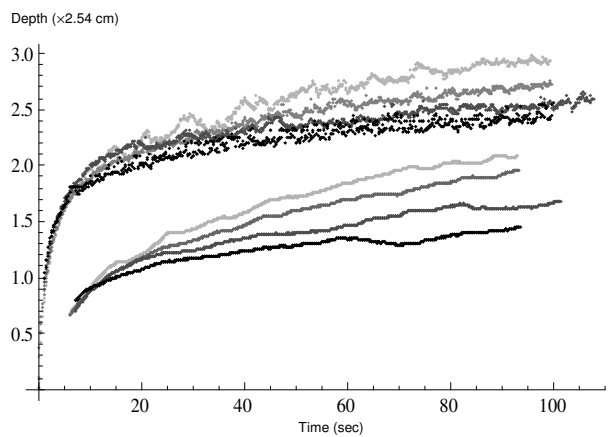}
  \caption{\label{craterdepths} Craters depths versus time for the same set of sands described in Fig.~\ref{craterwidths}.  The top set of curves shows the overall crater depths.  The bottom set of curves shows the inner crater depths, smoothed by a 101-point moving-average filter to remove the noise.}
\end{figure}

To investigate how cratering scales with particle size we used quartz ``construction sand'' that had been sieved to retain various size ranges.  For a 0.95 cm diameter pipe, 7.62 cm height above the sand, and a jet velocity of 34 m/s, the crater widths vs. time are shown in Fig.~\ref{craterwidths}.
The inner crater width quickly reaches a maximum whereas the outer crater continues to widen throughout the logarithmic period. This maximum width of the inner crater is found to be independent of particle size, whereas the outer crater widens faster for smaller particles.  Figure~\ref{craterdepths}
shows that unlike inner crater widths, the inner craters depths continued to grow throughout the logarithmic period and their rates were a function of particle size.  This was unexpected and lacks an explanation.  

\begin{figure}
  \includegraphics[angle=0,width=\columnwidth]{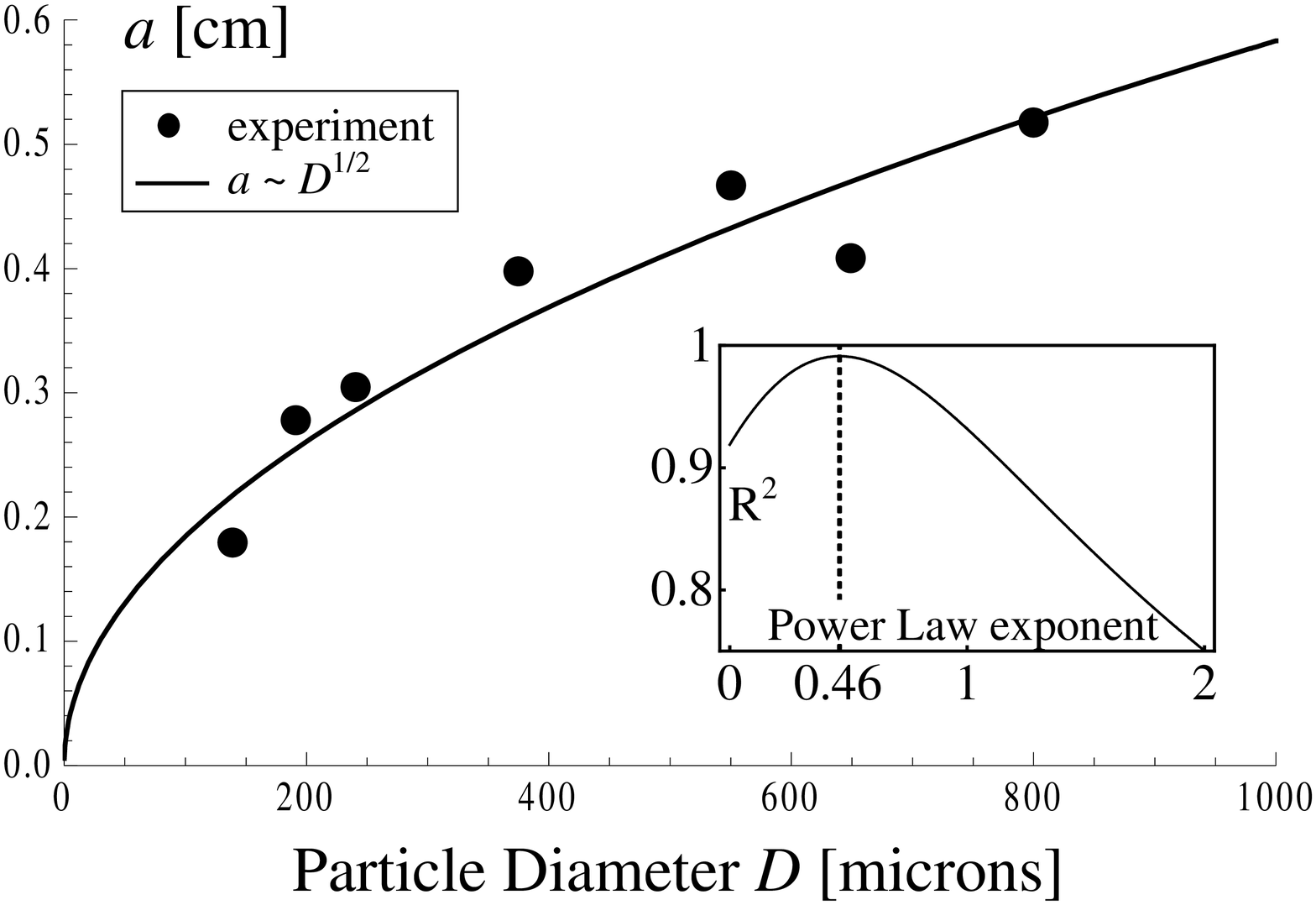}
  \caption{\label{aparam} The cratering length-scale $a$ vs. particle size $D$.  Inset:  the Coefficient of Determination $R^2$ for power laws of $D$ between 0 and 2 (best fit is 0.46 $\approx$ 1/2).}
\end{figure}

\begin{figure}
  \includegraphics[width=\columnwidth]{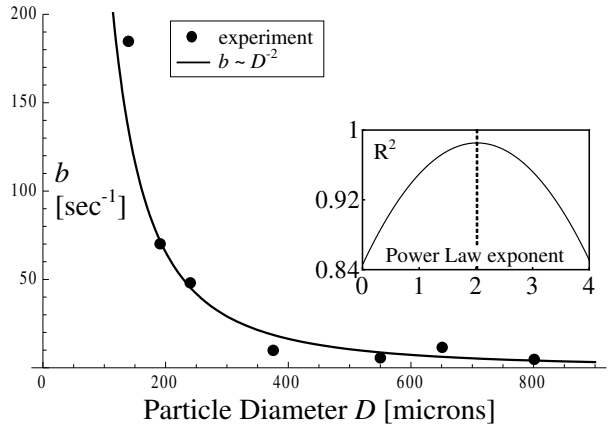}
  \caption{\label{bparam} The inverse time scale $b$ vs. particle size $D$.  Inset:  the Coefficient of Determination $R^2$ for power laws of $D$ between 0 and 4 (best fit is 2.02).}
\end{figure}

As described in \cite{metzger2}, straight lines were fitted to the measurements of overall depth vs. logarithmic time.  The slope of these straight lines is $a$, the dominant length-scale in the logarithmic period.  The intercept is $a\ln b$, where $b$ is the inverse of the dominant time-scale.  As reported in \cite{metzger2}, $a$ varies with the height of the jet exit plane above the soil, but surprisingly it is independent of jet diameter, jet velocity $v$, and gas density $\rho$.  However, $b$ varies with all these parameters and notably is proportional to the jet's dynamic pressure $\rho v^2$.  This is different than the scaling of the densimetric Froude number $Fr \sim \rho^{1/2} v$.  Here we report in Fig.~\ref{aparam}
and~\ref{bparam}
the scaling of $a$ and $b$ with respect to grain diameter $D$.  We find $a \sim D^{1/2}$ and $b \sim D^{-2}$.  We note that the volumetric erosion rate $a^3 b \sim \rho v^2 / D^{1/2}$ has the same scaling as the Froude number for $D$ but not for $\rho$ or $v$.

\subsection{Scaling With Gravity}

\begin{figure}
  \includegraphics[width=\columnwidth]{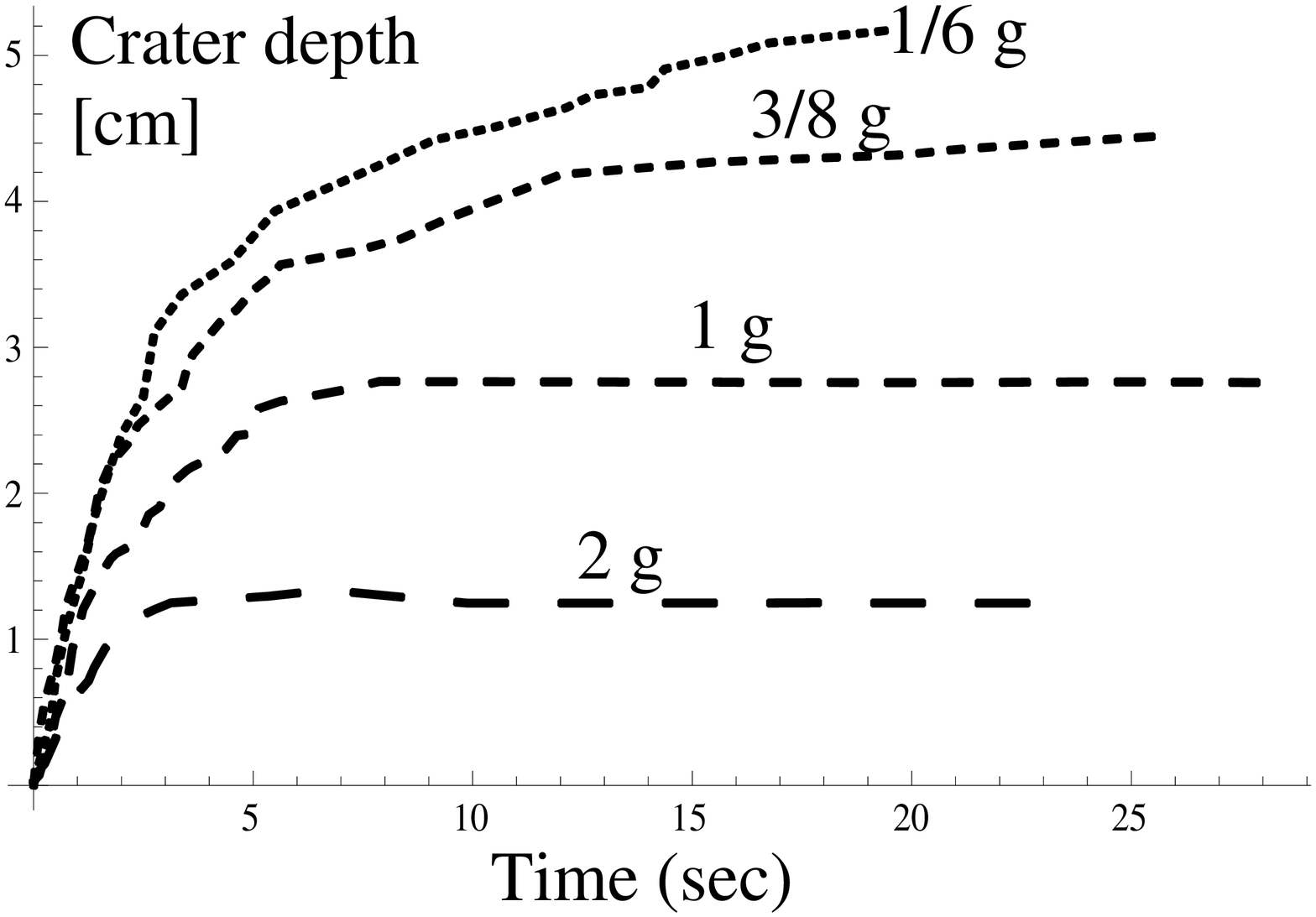}
  \caption{\label{gravitydepth} Overall craters depths versus time for simulated martian soil JSC-Mars-1A at four different gravity levels with all other parameters held constant.}
\end{figure}

\begin{figure}
  \includegraphics[width=\columnwidth]{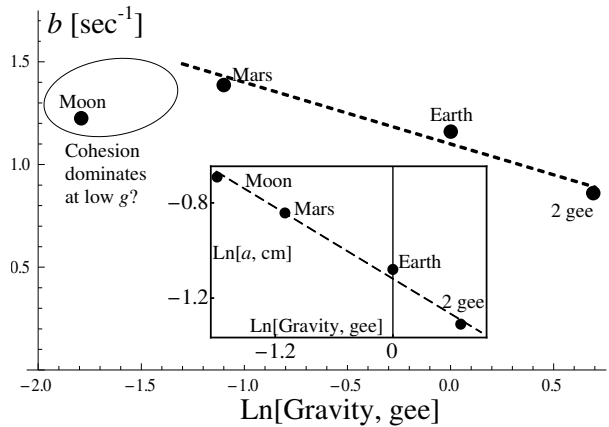}
  \caption{\label{abgravity} Cratering length-scale $b$ vs. the log of gravity.  The dashed line is a guide to the eye.  Inset:  log-log plot of cratering inverse-time scale $a$ vs. gravity.  The dashed line is $a \sim g^{-0.25}$.}
\end{figure}

We have performed similar experiments on an aircraft flying parabolic trajectories to effectively reduce gravity.  Seven different soils were used at a variety of jet velocities and four different gravities:  $g \approx 1/6$ gee (the Moon), 3/8 gee (Mars), 1 gee (Earth), and also 2 gee, which occurred during the ``pullouts'' of the aircraft between successive low-$g$ parabolas.  The low-$g$ experiments could not go longer than a single parabola because the craters caved-in during the periods of high-$g$.  Several hundred experiments were performed and the results are still being analyzed, but preliminarily we report one set of gravities for the simulated martian soil known as JSC-Mars-1A.  It is a volcanic soil with a broad particle size distribution from silt to coarse sand.  Figure~\ref{gravitydepth}
shows the depths versus time with the expected decrease in cratering rate at higher gravity levels.  The 1 gee and 2 gee experiments achieved asymptotic sizes within the duration of a parabola, with lesser gravity trending to deeper asymptotic craters.  Fig.~\ref{abgravity}
shows $a$ and $b$ for the logarithmic period at each gravity level.  In the 2 gee experiment the logarithmic period was very brief so $a$ and $b$ may have larger errors than for the other gravities.  Surprisingly neither parameter is strongly dependant on gravity, showing only $a \sim g^{-0.25}$ while the scaling of $b$ is inconclusive.  Cohesion probably dominates at low gravity and this is probably responsible for $b$ being non-monotonic with gravity.

\section{Bearing Capacity Failure beneath Supersonic Jets}

We also used solid rocket motors to test supersonic jets impinging on a deep sand bed against a viewing window.  When we placed layers of colored sand in the sandbox for some of these firings, the downward deflection of layers identified BCF as the dominant cratering mechanism.  We also noticed that in all our tests BCF continued until the crater was just slightly deeper than the length of the jet, and beyond that the much slower VE became dominant.  This led us to hypothesize that (within the parameter space of our experiments) the depth of BCF is controlled almost entirely by jet length rather than soil conditions, gravity, or other parameters.  

\begin{figure}
  \includegraphics[width=\columnwidth]{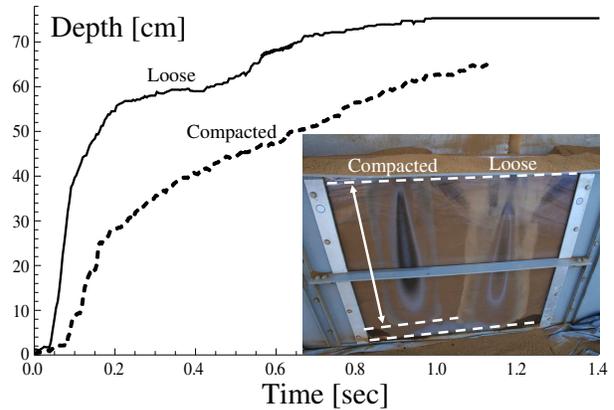}
  \caption{\label{supersonic} Depth of crater vs. time for two solid rocket motor firings into compacted and loosely deposited JSC-Mars-1A. Inset:  Scorch marks on the viewing window of the sand bed.}
\end{figure}

To test this hypothesis, we performed two successive firings into the sand bed filled with JSC-Mars-1A.  In the first test the soil was laid down in thin layers each vibrationally compacted.  In the second test the soil was poured in as loosely as possible.  The resulting soil density in a core extracted from the top 25 cm of the bed was measured to be four times higher in the vibrationally compacted state (1.04 vs. 0.25 g/$\textrm{cm}^3$), with three times higher shear strength (2.20 vs. 0.70 kg/$\textrm{cm}^2$ as measured by a Torvane shear tester), and seven times higher penetration resistance (9.70 vs. 1.40 kg/$\textrm{cm}^2$ as measured by handheld penetrometer).  For both motor firings we took high speed video and measured the depths of the craters versus time.  Also, the scorching of the viewing window provided witness marks for both the final crater depth and the jet core length in each test.  The results are shown in Fig.~\ref{supersonic}.
In both tests the BCF quickly achieved its maximum depth in about 1.5 seconds or less followed by the slower VE.  The final depth of the crater in loose soil was just slightly longer than in the compacted soil, and both were just slightly longer than the jets.  This confirmed our hypothesis:  in this region of the parameter space it was primarily the jet length and not the state of the soil that controlled the crater depth.  However, the BCF did achieve this depth faster in the loose soil.

\section{Conclusions}

Only small portions of the parameter space for jet-induced cratering have been sampled to-date.  The VE length $a$, time $b^{-1}$, and volumetric erosion $a^3 b$ scales in the logarithmic period are inconsistent with the Froude number.  Cohesion must be taken into account at low gravity.  BCF appears to be controlled primarily by the plume parameters (jet length).  Much work remains to understand the physics and scaling in the various flow regimes.

\bibliographystyle{aipproc}

\end{document}